\documentclass[structabstract]{aa}

\pdfoutput=1

\usepackage{graphicx}
\usepackage{txfonts}

\usepackage{natbib}
\bibpunct{(}{)}{;}{a}{}{,}

\begin{document}

\title{The baryonic Tully-Fisher Relation predicted\\by cold dark matter cosmogony}
\author{H.~Desmond\inst{}}

\institute{St. John's College, Oxford University, Oxford, OX1 3JP, UK\thanks{Currently: Physics Department, Stanford University, CA 94305-4060, USA} \email{harryd2@stanford.edu}}



\abstract {Providing a theoretical basis for the baryonic Tully-Fisher Relation (BTFR; baryonic mass vs rotational velocity in spiral galaxies) in the $\Lambda$CDM paradigm has proved problematic. Simple calculations suggest too low a slope and too high a scatter, and recent semi-analytic models and numerical galaxy simulations typically fail to reproduce one or more aspects of the relation. Furthermore, the assumptions underlying one model are often inconsistent with those behind another.} {This paper aims to develop a rigorous prediction for the BTFR in the context of $\Lambda$CDM, using only a priori expected effects and relations, a minimum of theoretical assumptions, and no free parameters. The robustness of the relation to changes in several key galactic parameters will be explored.} {I adopt a modular approach, taking each of the stand alone galaxy relations necessary for constructing the BTFR from up-to-date numerical simulations of dark halos. These relations -- and their expected scatter -- are used to describe model spirals with a range of masses, resulting in a band in the space of the BTFR that represents the current best guess for the $\Lambda$CDM prediction.} {Consistent treatment of expected $\Lambda$CDM effects goes a large way towards reconciling the ``na\"{\i}ve'' slope-3 $\Lambda$CDM prediction with the data, especially in the range $10^9 M_{{\odot}} < M_\mathrm{bar} < 10^{11} M_{\odot}$. The theoretical BTFR becomes significantly curved at $M_\mathrm{bar} > 10^{11} M_{\odot}$ but this is difficult to test observationally due to the scarcity of extremely high mass spirals. Low mass gas-rich galaxies appear to have systematically lower rotational velocity than the $\Lambda$CDM prediction, although the relation used to describe baryon mass fractions must be extrapolated in this regime. The fact that the BTFR slope derived here is significantly greater than in early predictions is a direct consequence of a corresponding increase in the expected sensitivity of baryon mass fraction to total halo mass.} {}


\maketitle

\section{Introduction}
\label{sec:Intro}

\subsection{The Tully-Fisher Relation in $\Lambda$CDM}
\label{sec:Prelim}

The Tully-Fisher Relation (TFR) was originally proposed as a correlation between 21 cm line width and optical luminosity in spiral galaxies~\citep{Tully_Fisher}, but it has subsequently become accepted that these are observational proxies for more fundamental physical properties, namely rotational velocity and stellar or baryonic mass~\citep[e.g.][]{Verheijen,McGaugh_BTFR}. The canonical approach towards deriving the TFR adopts the $\Lambda$CDM model of cosmology, in which the universe is flat and consists (by mass) of 4.6\% baryons, 23\% cold dark matter, and 73\% dark energy~\citep{Komatsu}. In this paradigm, visible galaxies are surrounded by roughly spherical halos of dark matter. Since it is total enclosed mass (baryonic plus dark) that determines a galaxy's rotational velocity at any particular radius, it is clear that the TFR describes a coupling between the baryons in a galaxy and the surrounding dark matter. The extent to which such a coupling is expected for cold, weakly interacting dark matter has recently been a hotly-debated subject, and consensus has not yet been reached~\citep[e.g.][]{Foreman_Scott,McGaugh_FSG_Reply}. The situation is complicated by the fact that many different forms for the TFR exist in the literature, using luminosities (at various wavelengths), stellar mass or baryonic mass as the dependent variable, and one of several different rotational velocity measures as the independent variable~\citep[e.g. see][]{Avila-Reese,Masters}.

The earliest $\Lambda$CDM predictions for the TFR started with the expected relation between the total mass of a galaxy and its halo, $M_\mathrm{vir}$, and the corresponding characteristic velocity dispersion $V_\mathrm{vir}$, assuming virial equilbrium up to a radius $R_\mathrm{vir}$. Within this, the galaxy's average mass is taken to be a fixed multiple of the background mass density of the universe; simple manipulations then yield $M_\mathrm{vir} \propto V_\mathrm{vir}^3$~\citep{White}. However, $M_\mathrm{vir}$ and $V_\mathrm{vir}$ cannot be measured directly: they are theoretical quantities pertaining to the putative dark matter halo. In order to make contact with the observed TFR, one must convert halo mass to observed mass or luminosity, and virial velocity to disk rotational velocity. The expectation was that the baryonic mass of a galaxy would be a fixed fraction of its total mass~\citep{White,MMW}, equal to the universal ratio of baryonic to total mass (now believed to be 0.17;~\citealt{Komatsu}). Assuming further that galaxies' rotational velocities are proportional to their virial velocities, one derives a prediction for the \emph{baryonic mass} TFR (BTFR): $M_\mathrm{bar} \propto V_\mathrm{rot}^3$. Such reasoning would provide a convenient explanation for a BTFR with a slope of 3 in log-log space. However, observational studies persistently find a slope greater than 3, typically in the range 3.5-4~\citep[e.g.~see][table~3]{GHASP}.

Several effects might be expected to invalidate the premises used to derive the slope-3 BTFR:

\begin{enumerate}

\item There is no a priori reason why a galaxy's rotational velocity should be proportional to its virial velocity. A better approximation would be to use the density profile of a galaxy and halo to compute the overall rotation curve, and to take the velocity at some characteristic point on this as the TFR's independent variable. The maximum rotation velocity will be used in this paper.\footnote{Selection effects arising from the use of the flat part of the rotation curve in some observational studies will be explored in Sect.~\ref{sec:Selection}.} For example, dark halos are commonly parametrized by the NFW density profile~\citep{NFW}:

\begin{equation}
\rho(r) = \rho_\mathrm{crit} \frac{\delta_0}{\frac{r}{r_s} (1+\frac{r}{r_s})^2},
\end{equation}
with $r_s$ the scale-length and $\delta_0$ a characteristic overdensity. $\delta_0$ may be written in terms of the concentration, $c = \frac{R_\mathrm{vir}}{r_s}$~\citep{MMW}:
\begin{equation}
\delta_0 = \frac{\Delta}{3} \frac{c^3}{\ln(1+c) - \frac{c}{1+c}}.
\end{equation}

$\Delta$ is the mean density of the galaxy, at the time of virialisation, in multiples of the critical density of the universe, $\rho_\mathrm{crit}$. It will be set to 200 in this paper.\footnote{$\Delta$ determines the notional extent of the dark halo ($R_\mathrm{vir}$), yet no consensus exists regarding exactly what value it should take (e.g. see~\citealt[][p.~2]{Zhao}). Halos do not have a well-defined edge, but rather merge continuously with the background mass density of the universe.}

The ratio of maximum rotational velocity to virial velocity depends on the concentration~\citep{Bullock}, which is found in N-body simulations to be a function of virial mass~\citep{Gao,Zhao}. After changing independent variable from virial velocity to maximum rotation velocity, this $M_\mathrm{vir}$-concentration relation is therefore expected to change the TFR's slope.

\item So far, no mention has been made of the collapse of the baryons in galaxies into disks, yet both variables in the BTFR are properties of these disks. In $\Lambda$CDM, the particles in a newly-virialised galaxy acquire angular momentum through the action of cosmological torques~\citep{Peebles}. Dissipative processes cause the baryons to lose energy, and hence fall towards the centre, and their rotation forces them into a flattened disk. This modifies the rotation curve of the galaxy. Furthermore, the distribution of dissipationless dark matter is affected by baryonic collapse in a process known as \emph{adiabatic contraction (AC)}. The shells that make up the dark halo are pulled gravitationally towards the disk, with their final positions fixed by conservation of specific angular momentum~\citep{Blumenthal,Gnedin_2004,Gnedin_2011}. This effect is also expected to modify the galaxy's rotation curve, and hence the velocity used in the TFR.

\item The proportion of a halo's mass that is in baryons is unlikely to be universal. In particular, baryons are expected to be more easily ejected (e.g. due to supernovae or stellar winds) from lower mass galaxies with shallower potential wells~\citep[e.g.][]{MassLoss_Gnedin,MassLoss_Hoeft}. This would make the baryon mass fraction a rising function of total (virial) mass. As well as directly determining the baryonic mass that constitutes the dependent variable of the BTFR, this $M_\mathrm{vir}$-$M_\mathrm{bar}$ relation governs the relative contributions of the disk and halo to the overall rotation curve, and hence affects the position of its maximum.

\end{enumerate}

Whilst the first two of these effects must be studied through numerical simulation of dark halos, the third is accessible observationally via weak lensing and satellite kinematics~\citep[e.g.][]{Mandelbaum,Conroy}. These techniques do indeed suggest that $M_\mathrm{bar} / M_\mathrm{vir}$ rises with $M_\mathrm{vir}$, but may only be used for galaxies with $M_\mathrm{star} \gtrsim 10^{9.5} M_{\odot}$~\citep[e.g. see][fig.~11]{Behroozi} and must therefore be extended to lower mass to make contact with the majority of TFR measurements. A recent technique for doing this is called Halo Abundance Matching (HAM;~\citealt{Puebla,Behroozi,Moster,Guo}). The premise of HAM is that the most massive halos harbour the most luminous galaxies, and thus galaxies observed in a particular survey may be uniquely assigned to dark halos formed in N-body simulations. In association with appropriate mass-to-light ratios for converting luminosity to stellar mass, and a relation between gas mass and stellar mass, the output of HAM is an $M_\mathrm{vir}$-$M_\mathrm{bar}$ relation for the galaxies in the survey. Encouragingly, this technique agrees well with the direct methods at high $M_\mathrm{star}$~\citep[e.g. see][fig. 11]{Behroozi}.

The way in which AC and the $M_\mathrm{vir}$-$M_\mathrm{bar}$ and $M_\mathrm{vir}$-concentration relations modify the ``na\"{\i}ve'' BTFR prediction is the primary concern of this paper. Before discussing the general approaches that have previously been employed in studies of this kind, it is important to describe an additional tension between observation and $\Lambda$CDM prediction that is exacerbated by the three effects listed above. It has been noted that the observed BTFR is very tight~\citep[e.g.][]{Verheijen,Trachternach,McGaugh_Complete}. Indeed, McGaugh argues in~\citet{McGaugh_Main} that the scatter of individual spirals from a power-law TFR may be entirely accounted for by observational uncertainties, suggesting that the theoretical relation has zero intrinsic scatter. However, several factors are expected to create significant scatter in the $\Lambda$CDM prediction, including variations in the density profiles of dark halos caused by different mass aggregation histories (manifest in scatter in the concentration of halos of a given mass; e.g. see~\citealt{Eisenstein,Jing}) and in baryon mass fractions~\citep[e.g][]{Behroozi,MMW}. These sources of scatter are described in more detail in Sect.~\ref{sec:Scatter}. McGaugh argues that such effects will inevitably make the predicted scatter irreparably larger than that in the data,  leading him to the conclusion that the BTFR is evidence against the $\Lambda$CDM paradigm itself. McGaugh proposes that the BTFR data are more consistent with the expectations of Modified Newtonian Dynamics (MOND), which implies a BTFR of the form $M_\mathrm{bar} \propto V_\mathrm{rot}^4$~\citep{Milgrom,McGaugh_Main}. This appears to be in agreement with recent observations of gas-rich spiral galaxies in addition to many observations of star-dominated spirals (\citealt[][hereafter MG12]{McGaugh_Complete}).

\subsection{The paths to an improved prediction}
\label{sec:Approaches}

Two main approaches exist for improving the ``na\"{\i}ve'' TFR prediction by taking into account the effects described above. The first is to perform a full numerical simulation of one or more spiral galaxies~\citep[e.g.][]{Piontek,Governato,Tissera}. Cosmological parameters and initial conditions are specified, and galaxies evolved stepwise through time according to general relativistic equations of motion and prescriptions for gas cooling, star formation, and stellar feedback. The baryonic or stellar masses of the resulting galaxies can then be measured in addition to their rotational velocities, enabling their positions on the TFR to be determined. In the future, this approach will likely provide the most complete and robust estimate for the TFR expected in $\Lambda$CDM.

Currently, however, such simulations face several problems that limit their usefulness, including insufficient resolution for successful modelling of galaxy structure, insufficient computing power for the simulation of a statistically significant number of spirals, and uncertainties in our theoretical understanding of dissipative baryonic physics. These problems are manifest in significant discrepencies between simulated spirals and both the observed TFR~\citep{Dutton_2010} and expected galaxy formation efficiencies~\citep{Guo}. Furthermore, this holistic approach risks masking the effects that individual correction factors have on the theoretical TFR: cosmological parameters are essentially fed into a `black box' describing a whole gamut of physical processes known to varying degrees of accuracy. Finally, approximations and uncertainties are introduced into the simulation that are not strictly required for prediction of (some versions of) the theoretical TFR. For example, star formation rates and thresholds are necessary for determining the stellar mass and luminosity of a galaxy -- and hence its position on the luminosity or stellar mass TFR -- but not the total baryonic mass. As another example, consider the $M_\mathrm{vir}$-$M_\mathrm{bar}$ relation. This may be predicted in numerical simulations from a prescription for the extent of baryon expulsion from a galaxy by supernovae and stellar winds, but may also be measured observationally (as documented in Sect.~\ref{sec:Prelim}). A disagreement would suggest a flaw in the simulations which would propagate into the predicted TFR, but the specific effect of this would not be readily visible. Using instead the \emph{observed} $M_\mathrm{vir}$-$M_\mathrm{bar}$ relation (or an extension thereof, such as that from HAM) would provide a more accurate prediction for the TFR, and would disentangle the result from any potential conflict between observed and simulated baryon mass fractions.

The alternative to complete cosmological simulation is semi-analytic modelling~\citep[see~e.g.][]{Hierarchical}. In this modular approach, the relationships between variables important for the TFR are determined empirically, where possible, and are otherwise given simple forms in accordance with the results of N-body simulations of dark halos. This allows the development of a TFR prediction with a minimum of assumptions and enables the individual effects of different components of the prediction to be investigated, thus helping to isolate the specific factors responsible for potential disagreement with observation. The disadvantage is that the various parameters used (those describing the $M_\mathrm{vir}$-$M_\mathrm{bar}$ and $M_\mathrm{vir}$-concentration relations, adiabatic contraction, and the halo density profile) are typically derived from different numerical simulations, which may use different initial conditions or halo virial parameters and therefore be marginally inconsistent. One of the first analyses of this nature was performed by~\citet[][hereafter M98]{MMW}, whose methodological framework provides a template for connecting the various pieces required for a prediction of the TFR. I adopt this template here.

When the ingredient relations are not believed to be well-known, free parameters are often introduced which may be tuned to produce a desired result. An example of such an approach is~\citet[][hereafter D07]{Dutton_2007}, where agreement with an observational TFR data set is optimised by searching through a high-dimensional parameter space for the region that minimises a $\chi^2$ goodness of fit estimator. This method is useful for telling us what the existence of the TFR implies for the properties of spiral galaxies. However, one cannot be said to have explained the TFR if one has simply selected parameter values that reproduce it. A basic tenet of my methodology (which is described in more detail in Sects.~\ref{sec:Philosophy} and~\ref{sec:Method}) is that the introduction of free parameters in the context of the TFR is not necessary and can in some circumstances produce misleading conclusions. The reasons for this, drawing examples from the methods and results of D07, are as follows:

\begin{enumerate}

\item It is necessary to assume some simple form for the relations in the first place in order to specify them with only a few parameters. However, these parametrizations themselves may lack physical motivation. For example, D07 use a power-law to describe the relation between $M_\mathrm{bar}$ and $M_\mathrm{vir}$ (a common approximation since the work of~\citealt{MMW}), but recent Halo Abundance Matching studies observe a turnover at high masses~\citep{Puebla,Moster}. These results will be inconsistent with those of D07 whatever parameter values are used.

\item D07 tune their parameters to produce agreement with a power-law that is taken to fully describe the TFR. Yet no reason for adopting a power-law parametrization is inherent in the data -- logically, such a form must instead be a property of a theoretical prediction. Indeed,~\citet[][hereafter TG11]{Trujillo-Gomez} find agreement between the data and a theoretical line that is significantly curved, suggesting that it is not necessary for a prediction to be a power-law in order to be acceptable. Furthermore, the slope and intercept of the power-law fit to the TFR may not be very well defined: it may be possible to fit the TFR almost equally well using two power-laws with totally different slopes and intercepts~\citep[e.g.][fig.~2]{Foreman_Scott}. An observed TFR comprising data points with uncertainties is not uniquely equivalent to a power-law with uncertainties in its slope and intercept.

\item Even if a power-law could fully describe the observed TFR, its parameters are obviously dependent on the data set used. There have been many studies into the TFR, all yielding slightly different results. Thus the results in D07 are strictly valid only for the particular observational data that they use to construct the TFR, intertwining theory and observation in a non-trivial way. In fact, different best fit slopes and intercepts can be given even for the \emph{same} data set, by the use of different fitting methods (e.g. see MG12, Sect. 2.6.1). Taking a particular power-law to be the correct description of the TFR requires the assumption that a particular fitting method is superior to all others. This is not the case: they are all just different approaches to an underdetermined statistical problem. It is not clear how changes to the parameters of the power-law TFR would affect the optimum values of the tunable parameters.

\item Although it will be possible to find the \emph{global} minimum of the $\chi^2$ value of the TFR fit, it may be that distant \emph{local} minima exist in the many-dimensional parameter space with almost as low a $\chi^2$ value. Given the uncertain nature of the observed TFR, these could provide equally good descriptions of disk galaxies and may, in fact, be better physically-motivated.

\item All of the free parameters are correlated in the context of the $\chi^2$ value that they produce, making it difficult to assign them unique and independent uncertainties.

\item The relationships obtained after tuning the free parameters lack physical significance because they have become divorced from the motivations that originally existed for introducing them. At the end, the results must be compared with independent measurements to check consistency, but, to the extent that such independent measurements exist, one may as well use them from the outset.

\end{enumerate}

\subsection{Two important methodological issues}

In this section I discuss two further methodological issues that influence the nature of the TFR and the way in which it is presented. By clarifying them here I intend to illuminate the approach to TFR prediction that I believe to be most constructive.

\subsubsection{Different versions of the Tully-Fisher Relation}
\label{sec:TFRs}

As alluded to in Sect.~\ref{sec:Prelim}, several different versions of the TFR exist. In particular, one may use luminosity, stellar mass or baryonic mass as the dependent variable, and the significance that one attributes to these different TFRs depends on the perspective brought to the problem. McGaugh, for example, believes the baryonic TFR to be fundamental because of its power-law nature and low intrinsic scatter -- and because it is such a relation that arises naturally in MOND (\citealt{McGaugh_Main}, MG12). Others (for example TG11 and~\citealt{Dutton_2010}) consider the TFR to be a consequence of complicated physical processes occuring during hierarchical galaxy formation. As such, the TFR is no more significant than the luminosity TFR (LTFR). Since it is luminosity that is directly observed, these authors take the latter as the touchstone for their $\Lambda$CDM models. (The baryonic TFR is clearly of secondary importance in TG11, where its intrinsic scatter is not even considered.) However, regardless of the MOND-$\Lambda$CDM debate, it seems intuitively reasonable that mass should be more fundamental than luminosity when it comes to a relation with rotational velocity; luminosity depends on the details of stellar populations, which are not relevant for determining how fast a galaxy spins. Even in $\Lambda$CDM, one would expect $M_\mathrm{vir}$ to be correlated more strongly with $M_\mathrm{bar}$ than $M_\mathrm{star}$, and hence the BTFR is again the more ``fundamental''.

If one starts with a $M_\mathrm{vir}$-$M_\mathrm{bar}$ relation yet wishes to construct a theoretical LTFR (as in D07, for example), two additional ingredients are required. First, a prescription for star formation is neccesary to convert baryonic mass to stellar mass. This is not required for predicting the BTFR --  where stellar (and gas) masses are measured directly to constitute the observational data points -- and unnecessarily introduces an extra degree of uncertainty. Put another way, comparison of the theoretical and observed LTFRs tests the hypothesis $\Lambda$CDM + star formation prescription, whilst comparing BTFRs tests $\Lambda$CDM alone. Disagreement with the expected LTFR may simply be due to errors in the star formation rate. The second required ingredient is a mass-to-light ratio to convert stellar mass to luminosity. Such a ratio is also required for consideration of the BTFR, but here for the observations as opposed to the theory. Reducing the number of assumptions and uncertain parametrizations that go into the theory allows it to be more directly compared with a range of observational data sets using different mass-to-light ratios.

Nevertheless, there do exist uncertainties inherent in the BTFR that are absent from the LTFR. If gas masses in spirals are not well known, the BTFR may be subject to large systematic uncertainties that render impossible a fair comparison with any $\Lambda$CDM prediction. In this vein,~\citet{Gnedin_IonisedGas} has argued that disagreement between BTFR observations and the $\Lambda$CDM prediction implies the existence of entirely unobserved baryons in spirals, which he believes to be in the form of warm ionised gas created by radiation from the inner stellar disk and the ionising cosmic background. Such claims are diametrically opposed to those of McGaugh, who argues that systematic errors such as missing gas mass should not be able to produce a tight BTFR correlation where none would otherwise exist. In other words, McGaugh uses his zero-intrinsic-scatter slope-4 BTFR (MG12, fig. 1) as evidence that all baryonic mass has been accounted for: it would be extremely unlikely for omission of one baryonic component to turn the messy relation predicted by $\Lambda$CDM into the precise power-law relation that is observed. McGaugh also cites observational studies which find insufficient ionised gas to significantly affect the BTFR~\citep[e.g.][]{Anderson}. In fact, it is argued in~\citet{McGaugh_Main} that the uncertainties involved in determining the gas mass (due to imperfectly-known distances and uncertainties in measured line fluxes) are dwarfed by those associated with the conversion of luminosity to stellar mass (which stem from stellar population modelling and the Initial Mass Function). This suggests that of all the variants of Tully-Fisher Relation, the BTFR should provide the cleanest test of any theory. In any case, it is the BTFR that we must investigate if we are to make contact with McGaugh's claim that the slope and scatter of this relation constitute evidence against $\Lambda$CDM.

One final point has great significance in the context of my work. Even if much ionised gas were to exist, this would affect neither the BTFR data points nor a theoretical prediction using baryon mass fractions obtained by Halo Abundance Matching. This is because the baryonic mass used in HAM is precisely the baryonic mass that we observe, and thus identical to that plotted in the BTFR. Use of an empirical $M_\mathrm{vir}$-$M_\mathrm{bar}$ relation from HAM renders the issue of BTFR prediction entirely independent of the missing mass problem (observed baryon mass fractions significantly different from the ``average'' cosmological value of 0.17). The conjecture of~\citet{Gnedin_IonisedGas} may shed light on the whereabouts of this missing mass, but has no ramifications for a BTFR built using HAM.

\subsubsection{Inclusion of elliptical galaxies}
\label{sec:Ellipticals}

Both TG11 and~\citet{Dutton_2010} introduce ellipticals into their TFR, thereby turning it into a more general ``luminosity-velocity'' relation. By enlarging the scope of the ``TFR'' in this way, they are able to check their predictions against more observational data and explore the differences in the properties of early- and late-type galaxies. However, it is not clear to what extent the direct comparison of spirals and ellipticals in this way is appropriate. $\sqrt{3} \sigma_{los}$ is used as an elliptical equivalent of rotational velocity, where $\sigma^2_{los}$ is the stellar velocity dispersion along the line of sight to the galaxy. This assumes virial equilibrium and an isotropic stellar velocity distribution; the (poorly-known) geometry of ellipticals can cause differences of up to 20\%~\citep{McGaugh_Wolf}. Further, one must be careful to measure $\sigma_{los}$ at a radius comparable to the one at which the rotational velocities of spiral galaxies are measured. The introduction of elliptical galaxies clearly creates several uncertainties that need not be of concern if one is dealing specifically with the TFR.

TG11 observe a difference in the position of spiral and elliptical galaxies in their mass-velocity plot (their fig. 11). As they themselves note, this is likely to be due to a dependence on galaxy morphology of the relationships which act as ingredients of the theoretical BTFR, for example the $M_\mathrm{vir}$-$M_\mathrm{bar}$ relation. No distinction is made between spirals and ellipticals in the Halo Abundance Matching that TG11 use -- both types of galaxy are assumed to have the same baryon mass fractions. Further, gas mass fractions are very much lower in ellipticals than they are in spirals (star formation has been completed in the former but is ongoing in the latter; e.g. see~\citealt{Puebla}, fig. 2), yet the empirical $M_\mathrm{star}$-$M_\mathrm{gas}$ relation used in TG11 for both types of galaxy~\citep[from][]{Baldry} is for spirals only. Thus their theoretical line strictly applies to neither spirals nor ellipticals on their own, but rather an amalgam of the two.\footnote{Incorrect gas mass fractions will have a far larger effect on the BTFR than the LTFR. Part of the reason why this discrepency is not considered to be very important by TG11 is the subordinance of the BTFR to the LTFR in their work.} Given the uncertainties introduced by ellipticals described in the paragraph above (and in light of the considerable amount of confusion that already exists surrounding the Tully-Fisher Relation itself) it seems sensible to remove ellipticals from the discussion. In the context of HAM, one should therefore match \emph{spiral} galaxies in the observational survey with \emph{simulated halos that are believed would harbour spirals} in reality. The latter may be identified by their environment and merger history, and matching of this type has, since the time of TG11's study, been performed in~\citet[][hereafter RP11]{Puebla}.

\subsection{Aim}
\label{sec:Philosophy}

In view of the above issues, my approach will be the following. I aim to construct a theoretical BTFR, taking account, as completely as possible, of all relevent effects that are seen in up-to-date numerical simulations of dark matter halos. I will use only parameters and relations that are expected a priori from analytic and numerical studies of dark halos, and will not include any tunable quantities. Where applicable, these relations will be specifically for spirals. In addition, I will propagate into the BTFR the expected scatter in all relations that I use, investigate the effect of adiabatic contraction, and determine the robustness of my results to changes in the halo density profile. The result will be the expected BTFR in the $\Lambda$CDM paradigm, which may be fairly compared to a range of observational data sets. Two such data sets will be used to qualitatively illustrate the degree of agreement between theory and observation. Finally, I will investigate the extent to which the prediction would be modified by modelling two selection effects in the observational data.

The structure of this paper is as follows. In Sect.~\ref{sec:Method}, I describe the technical details of my calculation of the predicted BTFR. Section~\ref{sec:Data} explores the issues involved in selecting appropriate observational data for comparison with theory, and describes the data sets that will be used. The results will be presented and discussed in Sect.~\ref{sec:Results}. Section~\ref{sec:Comp} contains a comparison of my findings with those of three recent studies in the literature. Finally, Sect.~\ref{sec:Conclusion} summarises my conclusions and suggests fruitful avenues for further work.

\section{Method}
\label{sec:Method}

The basic semi-analytic methodology employed in this work is adapted from M98 (Sect. 2). As the starting point, I take a spiral galaxy with a particular stellar mass. The $M_\mathrm{vir}$-$M_\mathrm{star}$ relation from the Halo Abundance Matching performed in RP11 (eq. 5) is then used to calculate the virial mass of the surrounding dark halo. The corresponding gas mass is calculated from the $M_\mathrm{star}$-$M_\mathrm{gas}$ relation described by RP11 eq. 2, in order to derive an analytic form for the HAM $M_\mathrm{vir}$-$M_\mathrm{bar}$ relation (RP11, fig. 5). The total halo mass is then used to derive the virial radius, virial velocity and scale-length of the halo (assumed for the moment to have an NFW density profile; see M98, eq. 2). Given these quantities, I find a self-consistent solution for the scale-length ($R_d$) of the spiral galaxy's disk (assumed exponential) that is formed by baryon collapse, and the total rotation curve. $R_d$ is found via calculation of the total energy of the halo and angular momentum of the disk (an integral over the rotation curve), and comparison with a dimensionless spin parameter $\lambda$ (see M98, eq. 9). This parameter is used because its distribution is well-constrained by numerical simulations, a fact that will become important in the discussion of scatter in Sect.~\ref{sec:Scatter}. Before this, $\lambda$ will be set equal to its mean value of 0.05. The response of the dark halo to disk formation follows the adiabatic contraction prescription laid out in~\citet{Gnedin_2011}. $R_d$ depends on the concentration of the halo, which will be derived from $M_\mathrm{vir}$ according to the results of~\citet{Zhao}, fig. 16.

Starting with an estimate for $R_d$, the adiabatic contraction equations may be solved\footnote{\citet{Gnedin_2011} demonstrates that no single set of parameter values reliably specifies the effect of adiabatic contraction in all cases, and recommends considering a range. All plausible values are found to produce almost identical TFRs, so for simplicity I will use $A_0$ = 1.6, w = 0.8~\citep[see][eq.~4]{Gnedin_2011} throughout this work.} using the Newton-Raphson method to determine the final density profile of the dark halo and thence the rotation curve (a sum in quadrature of disk and halo contributions). The latter may be used to find a new estimate of the disk's angular momentum, and hence of $R_d$. This updated scale-length is then adopted and the process iterated until the $R_d$ value generated from the rotation curve is identical to the value that produces that curve. (My convergence requirement is that $f_r$ [see M98, eq. 29] change by less than 0.1\% in the final iteration; reducing this is found to make a negligible difference to the results.) The self-consistent rotation curve is then sampled at 1000 radii logarithmically evenly spaced between $0.0001 \times R_\mathrm{vir}$ and $0.4 \times R_\mathrm{vir}$, and 100 radii between $0.4 \times R_\mathrm{vir}$ and $R_\mathrm{vir}$, in order to determine the maximum rotation velocity of a model galaxy with this mass. Finally, this procedure is repeated for 200 stellar masses logarithmically evenly spaced in the range $10^5 M_{\odot} \leq M_\mathrm{star} \leq 10^{12} M_{\odot}$, allowing the predicted BTFR (maximum rotational velocity vs baryonic mass) to be plotted.

\section{Observational data}
\label{sec:Data}

There have been numerous studies measuring the baryonic mass and rotational velocity of actual spiral galaxies, all producing results that vary to a greater or lesser extent from one to another. Reasons for this include (but are not limited to) differences in the following: 1) Wavelength used for luminosity measurements; 2) Mass-to-light ratio used for converting luminosity to stellar mass; 3) Method for accounting for mass of atomic, molecular and ionised gas; 4) Method for measuring rotational velocity (e.g. using line width or a resolved rotation curve); 5) Radius at which the rotational velocity is measured (for rotation curve studies); 6) Type of galaxy survey; 7) Criteria for deciding exactly which spirals should be included. The precise effects of variations in these factors are poorly-understood, but are certain to produce non-trivial systematic shifts in the resulting BTFRs (e.g. see~\citealt{Foreman_Scott}, p. 3 and refs. therein). The nature of the observational data used for comparison with a BTFR prediction is therefore crucial for determining whether or not the prediction appears successful. This makes imperative the preliminary establishment of rigorous criteria for selecting observational data sets which have maximum compatibility with a given theoretical model.

The first step in this endeavour is to investigate the approximations and assumptions that go into the theory. The Halo Abundance Matching used to derive the $M_\mathrm{vir}$-$M_\mathrm{bar}$ relation contains both a prescription for calculating stellar mass from luminosity, and gas mass from stellar mass. Thus the resulting BTFR prediction should be compared with observational data employing the same mass-to-light ratios and techniques for gas mass measurement. In addition, I use the maximum rotational velocity to describe my model galaxies; observed galaxies should use a similar rotation measure. By using only observations with identical assumptions to a theoretical prediction, unknown systematic effects associated with variation in the parameters listed above will be eliminated. As long as the observations are consistent \emph{among each other}, cuts may be applied to the model galaxies to enhance compatibility with the data. In principle (if the observational uncertainties were sufficiently well-known) a hypothesis test such as $\chi^2$ could then provide a quantitative measure of the extent of agreement between theory and observation.

From this perspective, it may be counterproductive to bin together the results of many different galaxy surveys without regard of the different assumptions and approximations on which they are based (as done for example in TG11, fig. 11). The advantage of comparison with a large quantity of observational data is offset by a reduction in quality in the sense that the observational details of the surveys are lost. Further undesirable features of this method of data presentation are the following:

\begin{enumerate}

\item The error bars show the \emph{scatter} of the data points in each bin around their mean. Information concerning the \emph{uncertainties} on the individual measurements is entirely lost, making it impossible to judge whether or not the points could actually be consistent with the theoretical line.

\item It is not known how many data points are in each bin. Some bins may contain many more than the others, and should therefore be weighted more heavily when comparing to the theoretical line. This cannot be taken into account in a visual assessment.

\item The mass and velocity range of the data points in each bin is unknown -- the standard deviation is the only measure of dispersion that is retained.

\item Projecting each data point onto one of a discrete set of rotational velocity values causes loss of accuracy.

\end{enumerate}

I will use two observational data sets for comparison with the BTFR prediction generated by the methodology of Sect.~\ref{sec:Method}. I do not intend this to constitute a definitive test of theory, but rather a rough qualitative assessment of my results: a comprehensive literature review to find all data sets consistent with the principles above is beyond the scope of this paper. Indeed, my intention is to produce a prediction that includes a minimum of theoretical assumptions and may therefore be compared to a range of data sets as others see fit (and indeed as BTFR observations improve and proliferate). In the next two sub-sections I will describe the data that I use, and my reasons for selecting them.

\subsection{Gas-rich spirals}
\label{sec:McGaugh}

As discussed in Sect.~\ref{sec:TFRs}, a major (if not the dominant) source of uncertainty in BTFR measurements is the mass-to-light ratio used to determine $M_\mathrm{star}$. Thus it makes sense to prioritise gas- as opposed to star-dominated spirals, for which the relative contribution of the uncertainty in $M_\mathrm{star}$ to the error budget is low. MG12 presents a compilation of 47 spirals which have $M_\mathrm{gas} > M_\mathrm{star}$ and which moreover satisfy several quality criteria including a resolved rotation curve and consistent optical and HI inclinations. McGaugh uses the flat part of the rotation curve as the rotational velocity measure -- this can be approximately accounted for in my modelling by discarding galaxies with non-flat rotation curves (see Sect.~\ref{sec:Selection}). A further discrepancy between McGaugh's approximations and those of Sect.~\ref{sec:Method} is that RP11's Halo Abundance Matching uses mass-to-light ratios derived from the~\citet{YMB09} stellar mass function, whilst McGaugh assumes a Portinari population synthesis model and Kroupa Initial Mass Function. However, the gas-dominated nature of the MG12 galaxies renders them relatively insensitive to variations in the mass-to-light ratio.

\subsection{The GHASP survey}
\label{sec:GHASP}

No gas-dominated spirals exist with $M_\mathrm{bar} \gtrsim 10^{10.5} M_\mathrm{\odot}$. To extend the BTFR to higher-mass (e.g. to better constrain its slope), the data from~\citet{GHASP} will be used. This is derived from a homogeneous galaxy survey (GHASP) undertaken using a scanning Fabry-Perot interferometer in France. Besides homogeneity, the GHASP survey offers the advantages of a completely resolved 2D velocity field (enabling rotation curves to be determined without uncertainties concerning position angle or inclination), observation of close galaxies (minimising distance uncertainties), and the adoption of $V_\mathrm{max}$ as the rotational velocity measure (in accordance with my theoretical model). Mass-to-light ratios are obtained using a Bell population synthesis model, which is shown in RP11's fig. 1 to produce a similar galaxy stellar mass function to the~\citet{YMB09} results used in the Halo Abundance Matching.

\section{Results and discussion}
\label{sec:Results}

\subsection{The baryonic Tully-Fisher Relation predicted by $\Lambda$CDM}
\label{sec:First}

In this subsection, scatter in the BTFR prediction is entirely suppressed. Fig.~\ref{fig:Lines} compares three theoretical BTFRs with the observational data described in Sect.~\ref{sec:Data} (red\footnote{One of the galaxies from MG12 (DDO 210) was removed due to its extremely low baryonic mass.} and blue points).\footnote{The green data points are from TG11. A comparison with these will be presented in Sect.~\ref{sec:TG_Comp}.} In each case, an NFW halo density profile is assumed. The black line uses the $M_\mathrm{vir}$-$M_\mathrm{bar}$ relation for spirals only (RP11 table 1, column 3) and includes the effect of adiabatic contraction, and is therefore the primary result of this subsection. This prediction is contrasted with two others, one for which adiabatic contraction is switched off (i.e. halo unaffected by disk formation; cyan) and another which uses the $M_\mathrm{vir}$-$M_\mathrm{bar}$ relation for ellipticals (magenta). This takes $M_\mathrm{vir}$ to be calculated from $M_\mathrm{star}$ using the results in column of 4 RP11's table 1, and $M_\mathrm{gas}$ from the red line in RP11, fig. 2, and is displayed to illustrate the extent to which contamination of the HAM $M_\mathrm{vir}$-$M_\mathrm{bar}$ relation with elliptical galaxies may be expected to distort the BTFR (see also Sect.~\ref{sec:TG_Comp}). Some observations from this plot:

\begin{figure}
  \centering
  \includegraphics[width=0.5\textwidth]{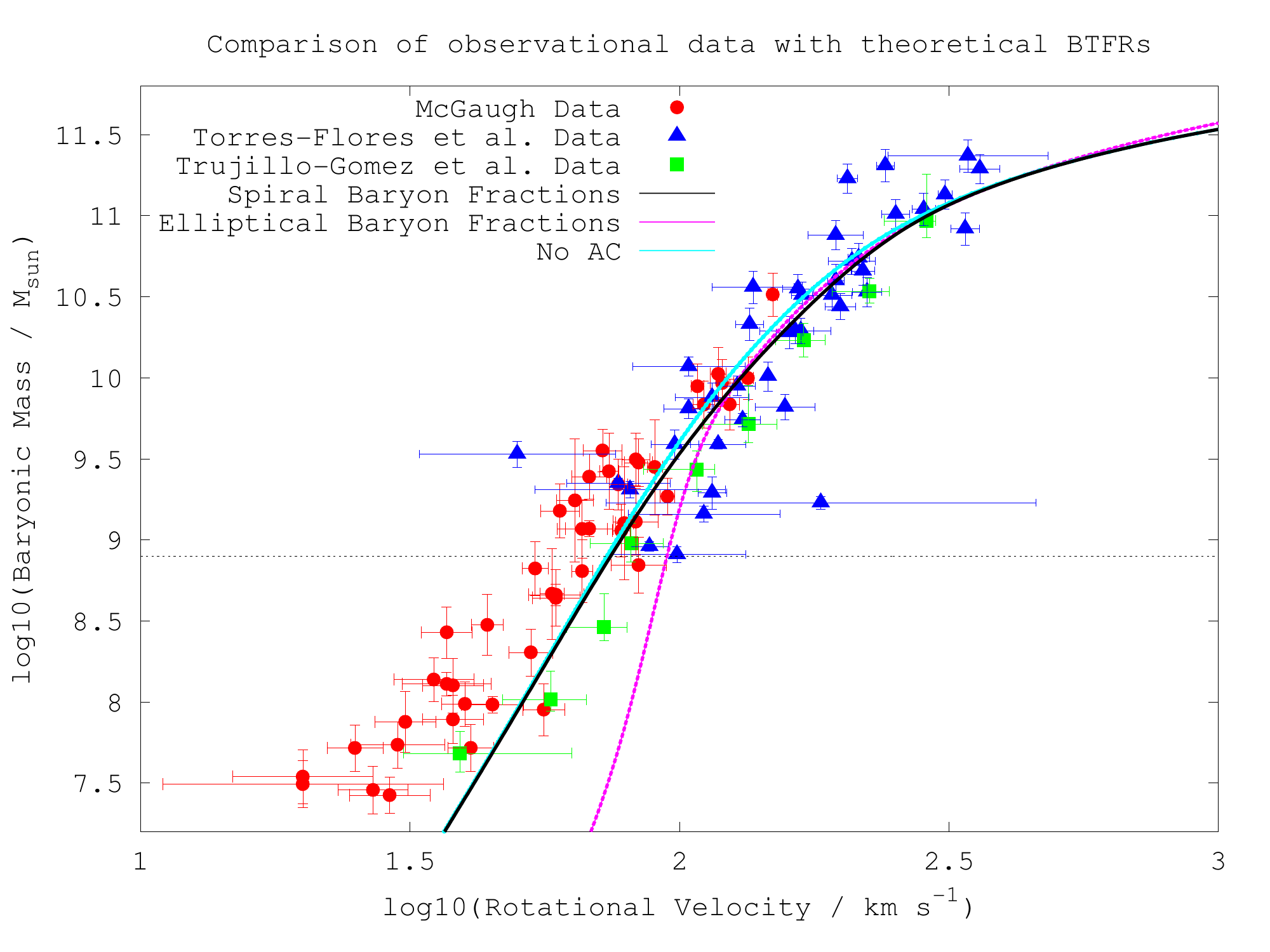}
  \caption{The black (magenta) line uses the $M_\mathrm{vir}$-$M_\mathrm{star}$ and $M_\mathrm{star}$-$M_\mathrm{gas}$ relations for spirals (ellipticals) -- see RP11, eq. 5 and fig. 2. The cyan line is the analogue of the black line but with AC switched off. Red data points are from~\protect\cite{McGaugh_Complete}, blue from~\protect\cite{GHASP} and green from~\protect\cite{Trujillo-Gomez}. The $M_\mathrm{vir}$-$M_\mathrm{bar}$ relation must be extrapolated below the dotted horizontal line.}
  \label{fig:Lines}
\end{figure}

\begin{enumerate}

\item Each line is almost a power-law up to $M_\mathrm{bar} \approx 10^{10.5} M_{\odot}$, beyond which it becomes much shallower. This is clearly due to the nature of the $M_\mathrm{vir}$-$M_\mathrm{bar}$ relation, in which a similar break occurs (RP11, fig. 5, top right panel). At high virial masses, $M_\mathrm{bar}$ rises very slowly with $M_\mathrm{vir}$. Thus the increase in rotational velocity is large relative to that in baryonic mass and the BTFR flattens out.

\item The curvature becomes very obvious around  $M_\mathrm{bar} \approx 10^{11} M_{\odot}$. However, this is approximately where the observational data ends. The data does not, therefore, provide compelling evidence for either a continuation of the power-law from lower masses (as expected for example in MOND) or for the decrease in slope predicted here. Observations of higher-mass spirals would be needed to decide between these two possibilities.

\item There is quite good agreement with the entire GHASP sample. However, the predictions lie systematically below the data for the gas-dominated spirals. In this respect, the black line gives a better fit than the magenta line, suggesting that a failure to discriminate between spirals and ellipticals will reduce agreement with the data at low baryonic mass (as might be expected from the fact that the observations are specifically of spirals).\footnote{This issue will be discussed further in Sect.~\ref{sec:TG_Comp}.} It is crucial to note, however, that the $M_\mathrm{vir}$-$M_\mathrm{bar}$ relation from RP11 must be extrapolated for $M_\mathrm{bar} \lesssim 10^{8.9} M_{\odot}$ (see RP11, fig. 5), increasing the uncertainty in the theoretical prediction in this region. This threshold is shown by the horizontal dotted line.

\item The effect of AC is to slightly reduce $M_\mathrm{bar}$ for spirals with $10^2~\mathrm{km~s^{-1}} \lesssim V_\mathrm{rot} \lesssim 10^{2.5}~\mathrm{km~s^{-1}}$. This may marginally improve agreement with the observational data.

\end{enumerate}

\begin{figure}
  \centering
  \includegraphics[width=0.5\textwidth]{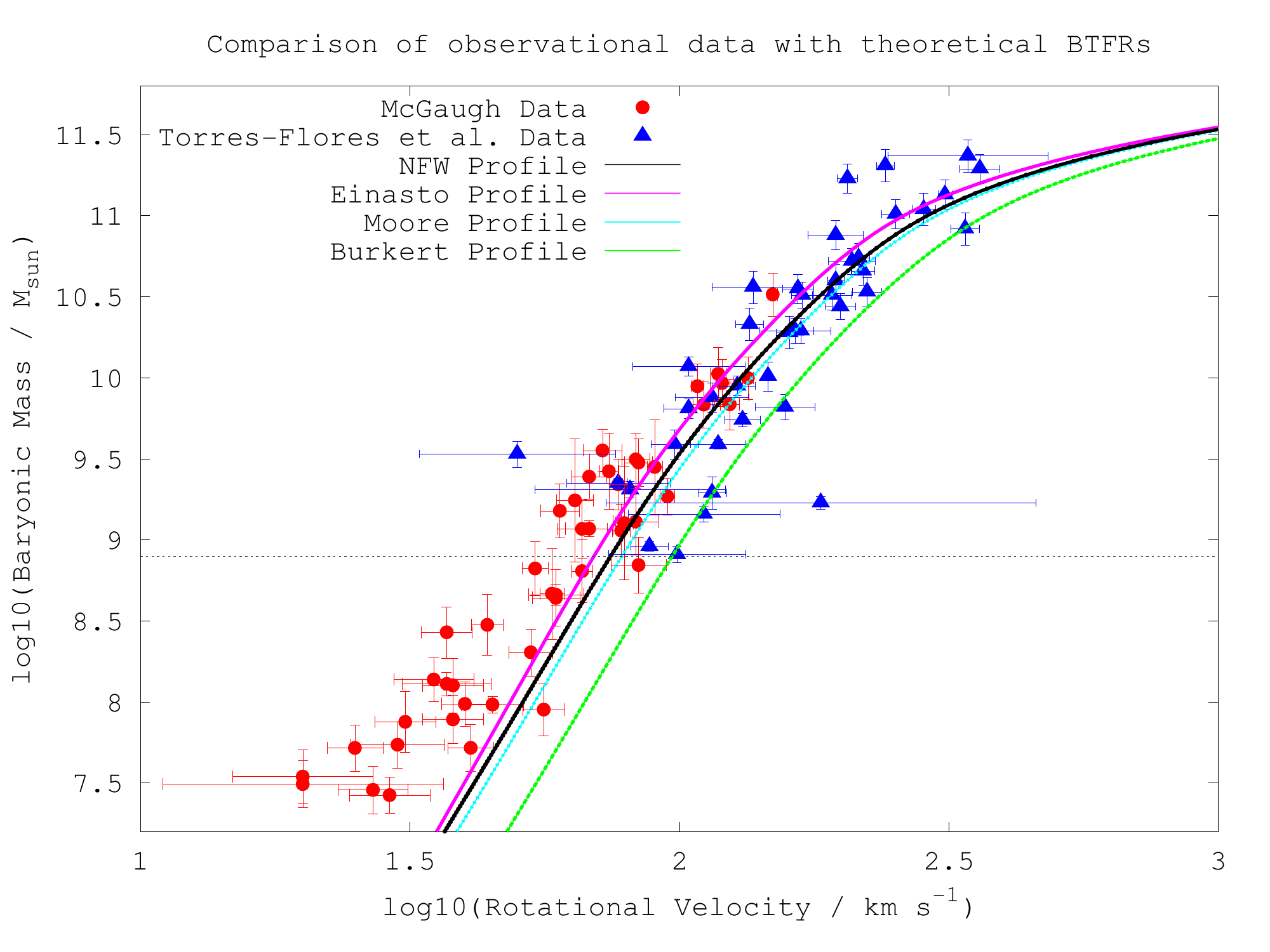}
  \caption{Data points as in Fig.~\ref{fig:Lines}. See text for further details.}
  \label{fig:Profiles}
\end{figure}

Fig.~\ref{fig:Profiles} shows the results when the density profile of the model halos is switched from NFW to Einasto~\citep{Gao}, Moore~\citep{Moore}, or Burkert~\citep{Burkert}. To construct these lines, eqs. 19, 21 and 23 of M98 were modified according to the change in $\rho(r)$. For the Moore and Burkert profiles, an identical $M_\mathrm{vir}$-concentration relation to the one used for NFW was adopted.\footnote{Concentration depends on halo scale-length, which strictly is only defined for the NFW profile. For alternatives, the scale-length will be taken as the radius at which the density becomes proportional to $r^{-2}$.} Concentrations for the Einasto profile were derived from $M_\mathrm{vir}$ using eq. 6 of~\citet{Gao}, a study working explicitly with this density profile. The Einasto shape parameter $\alpha$ was calculated using~\citet{Gao}, fig. 2. The curvature in the BTFR prediction is largest using the Einasto profile and smallest using Burkert, but, in general, changing the density profile has relatively little effect. Whilst the Einasto result fits the data somewhat better at the highest and lowest baryonic masses, the NFW line appears near-optimal in the range $10^9 M_{\odot} < M_\mathrm{bar} < 10^{11} M_{\odot}$.

\subsection{Intrinsic scatter}
\label{sec:Scatter}

Three sources of scatter in the theoretical BTFR are variability in halo spin, concentration and baryon mass fraction. The spin parameter $\lambda$ is determined by cosmological torques and inter-galactic tidal interactions~\citep{Eisenstein}, and therefore varies with environment. Numerical simulations find $\lambda$ following a log-normal probability distribution with mean 0.05 and standard deviation 0.5~\citep{MMW}. Galaxies of the same mass may have different concentrations (differently shaped density profiles) due to differences in their merger history and hence epoch of virialisation. Although no uncertainty is quoted in the source of the $M_\mathrm{vir}$-concentration relation used here~\citep{Zhao}, other studies have indicated that the concentration follows a log-normal distribution with standard deviation roughly independent of $M_\mathrm{vir}$ and equal to 0.18~\citep{Jing,Bullock}. Scatter in the HAM $M_\mathrm{vir}$-$M_\mathrm{bar}$ relation comes from a variety of sources (RP11, Sect. 3.1.1; see also~\citealt{Behroozi}). RP11 quote the scatter in $\log M_\mathrm{bar}$ as 0.23 dex at all virial masses.

To model the effect that these sources of scatter have on the intrinsic scatter of the BTFR, I make the following modifications to the methods used to generate the black line of Fig.~\ref{fig:Lines}: 1) The number of $M_\mathrm{star}$ values is increased from 200 to 500. 2) For each $M_\mathrm{star}$, 500 different values of $\lambda$, $M_\mathrm{bar}$ and concentration are randomly drawn from their respective probability distributions. The mean of the log-normal $M_\mathrm{bar}$ and concentration distributions are the values of these parameters used in Sect.~\ref{sec:First}. 3) A 150$\times$150 element grid is constructed spanning the range $10^6~M_{\odot} < M_\mathrm{bar} < 10^{12}~M_{\odot}$, $10^{0.7}~\mathrm{km~s^{-1}} < V_\mathrm{max} < 10^{2.7}~\mathrm{km~s^{-1}}$. For each of the 500$\times$500 = 250\,000 input parameter sets, $M_\mathrm{bar}$ and $V_\mathrm{max}$ are calculated and identified with a particular element of this grid. The final number of points in each element is then outputted and colour-coded for comparison with the observational data.

\begin{figure}
  \centering
  \includegraphics[width=0.5\textwidth]{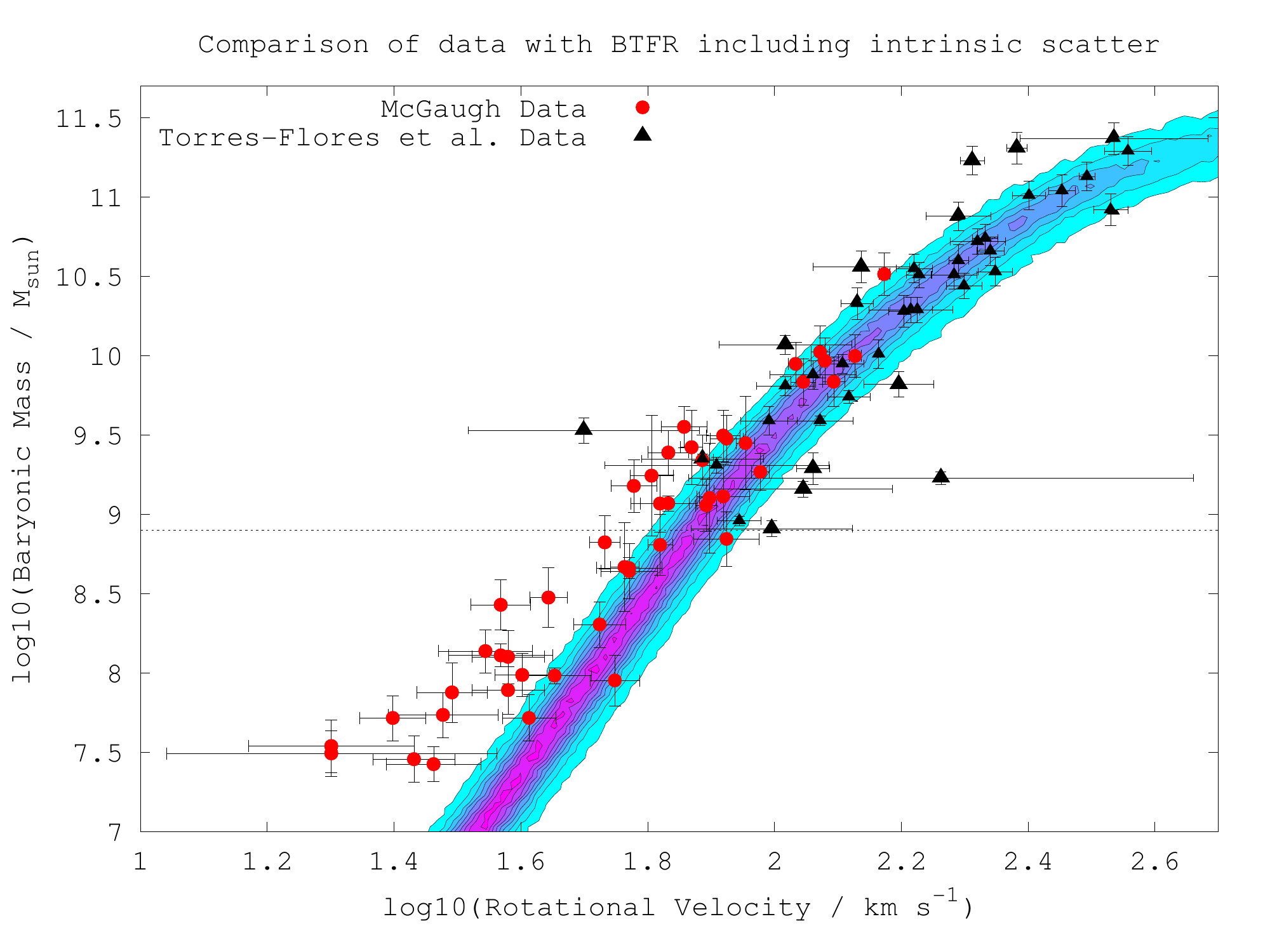}
  \caption{Dependence of model galaxy count on region of the $M_\mathrm{bar}$-$V_\mathrm{max}$ plane, following the procedure of Sect.~\ref{sec:Scatter}. The contours are at 10, 50, 100, 150, 200, 250, 300, 350, and 400.}
  \label{fig:Scatter}
\end{figure}

The resulting contour plot constitutes Fig.~\ref{fig:Scatter}, and shows the band in which observational points are predicted to lie, taking into account the expected intrinsic scatter. A little under 250\,000 model galaxies lie within the mass and velocity ranges shown in this figure, and a grid element is coloured if it contains at least 10 galaxy points. Thus it should be rare, in the $\Lambda$CDM model presented here, for galaxies to lie outside the coloured band. As with the black line in Fig.~\ref{fig:Lines}, there is good agreement with the data at $M_\mathrm{bar} \gtrsim 10^9 M_{\odot}$ and poor agreement below (where the HAM $M_\mathrm{vir}$-$M_\mathrm{bar}$ relation must be extrapolated). The amount of intrinsic scatter looks to be in reasonable accord with that in the data.

\begin{figure}
  \centering
  \includegraphics[width=0.5\textwidth]{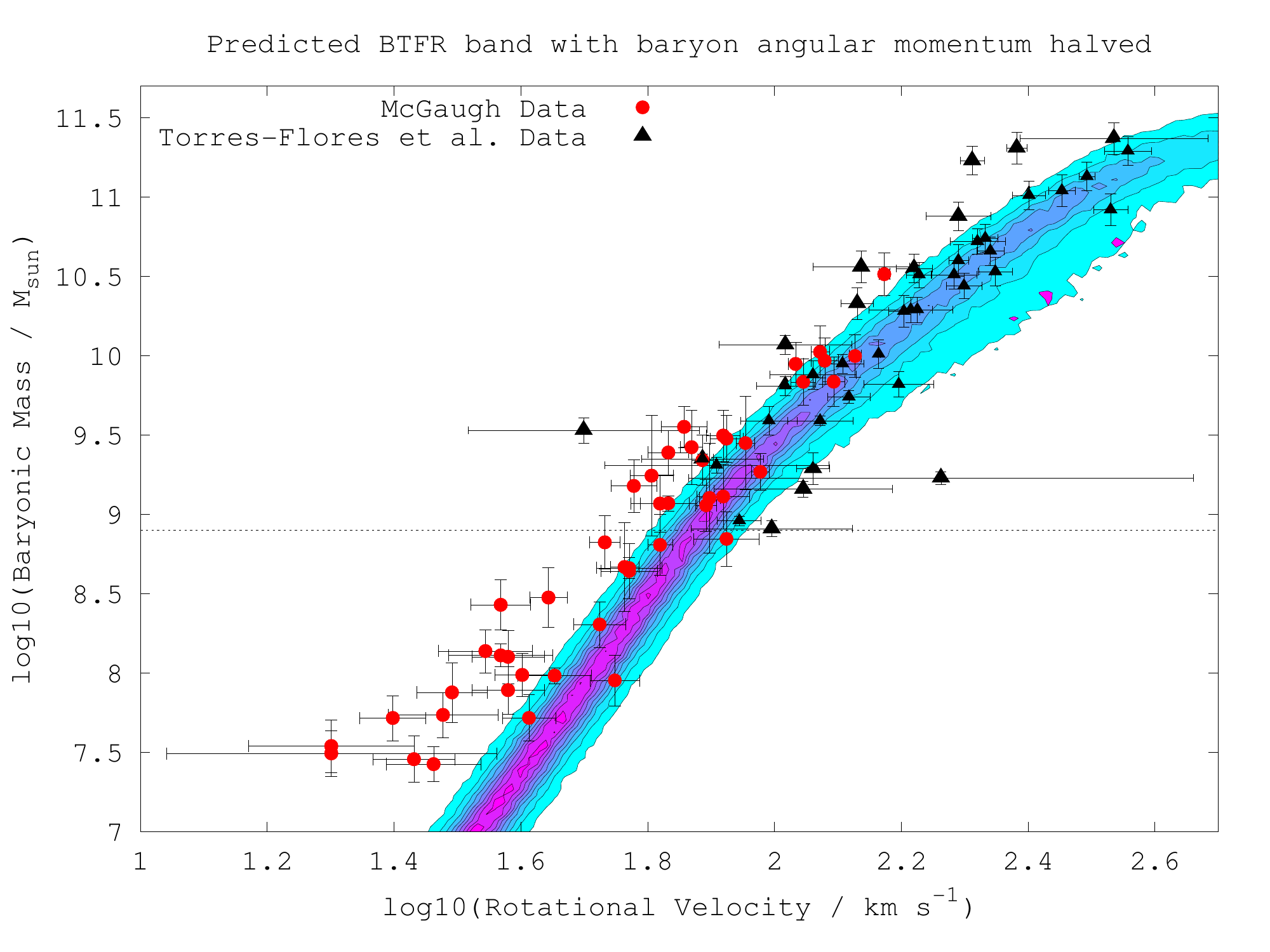}
  \caption{As Fig.~\ref{fig:Scatter}, but with the baryon angular momentum fractions of all model galaxies halved.}
  \label{fig:Scatter_jd}
\end{figure}

An additional quantity required for calculation of the maximum rotation velocity of each model galaxy is the fraction of the galaxy's total angular momentum that belongs to the baryons ($j_d$; see M98, eq. 8). Thus far, $j_d$ has been set equal to $M_\mathrm{bar}$ / $M_\mathrm{vir}$, the fiducial assumption of M98. However, this paper also cites evidence from numerical simulations (in Sect. 2.2) that this may overestimate $j_d$. To test the robustness of my results to a decrease in baryon angular momentum fraction, I plot in Fig~\ref{fig:Scatter_jd} the analogue of Fig.~\ref{fig:Scatter} but with all $j_d$ values halved. The effect is seen to be a slight downward shift of the predicted band at $M_\mathrm{bar} \gtrsim 10^{9.5} M_{\odot}$ (i.e. increased curvature) in addition to a small increase in scatter in this region. This decreases agreement with the high-mass data points, but to a sufficiently marginal extent for plausible changes in $j_d$ to be considered inconsequential.

\subsection{Selection effects}
\label{sec:Selection}

In Sect.~\ref{sec:Scatter}, all allowed halo spins, concentrations and baryon mass fractions were used to construct model galaxies, and contributed to Fig.~\ref{fig:Scatter}. However, it is unlikely that all such parameter values would yield spiral galaxies whose properties astronomers would measure and plot on the BTFR. Reasons for this fall in one of two categories: 1) Halos with extreme properties may be unstable and hence never form disk galaxies; 2) Astronomers typically require galaxies to fulfill certain selection criteria to be included in the TFR, with the intention of minimizing systematic uncertainties and ensuring some degree of consistency within the sample. A proper comparison of a theoretical BTFR with observational data should limit the input parameter space to the regions which generate stable spiral galaxies with properties passing all selection requirements of the data (see also Sect.~\ref{sec:Data}). In this section, I explore two such selection effects, one from each of the categories listed above.

It is theoretically expected that low $\lambda$ will make a spiral galaxy prone to the development of a bar instability, potentially causing either the total disruption of the disk or its transformation into an irregular galaxy which would not be plotted on the TFR~\citep[see e.g.][]{Christodoulou}. Removing the unstable model galaxies from the contour plot may therefore boost compatibility with the observational data. A simple prescription for the stability threshold is given by M98 (Sect. 3.2) in terms of the mass and scale-length of the disk, and maximum rotational velocity of the halo. Since these quantites are calculated for each model galaxy as part of the BTFR calculation, removing unstable galaxies in this way is straightforward.

\begin{figure}
  \centering
  \includegraphics[width=0.5\textwidth]{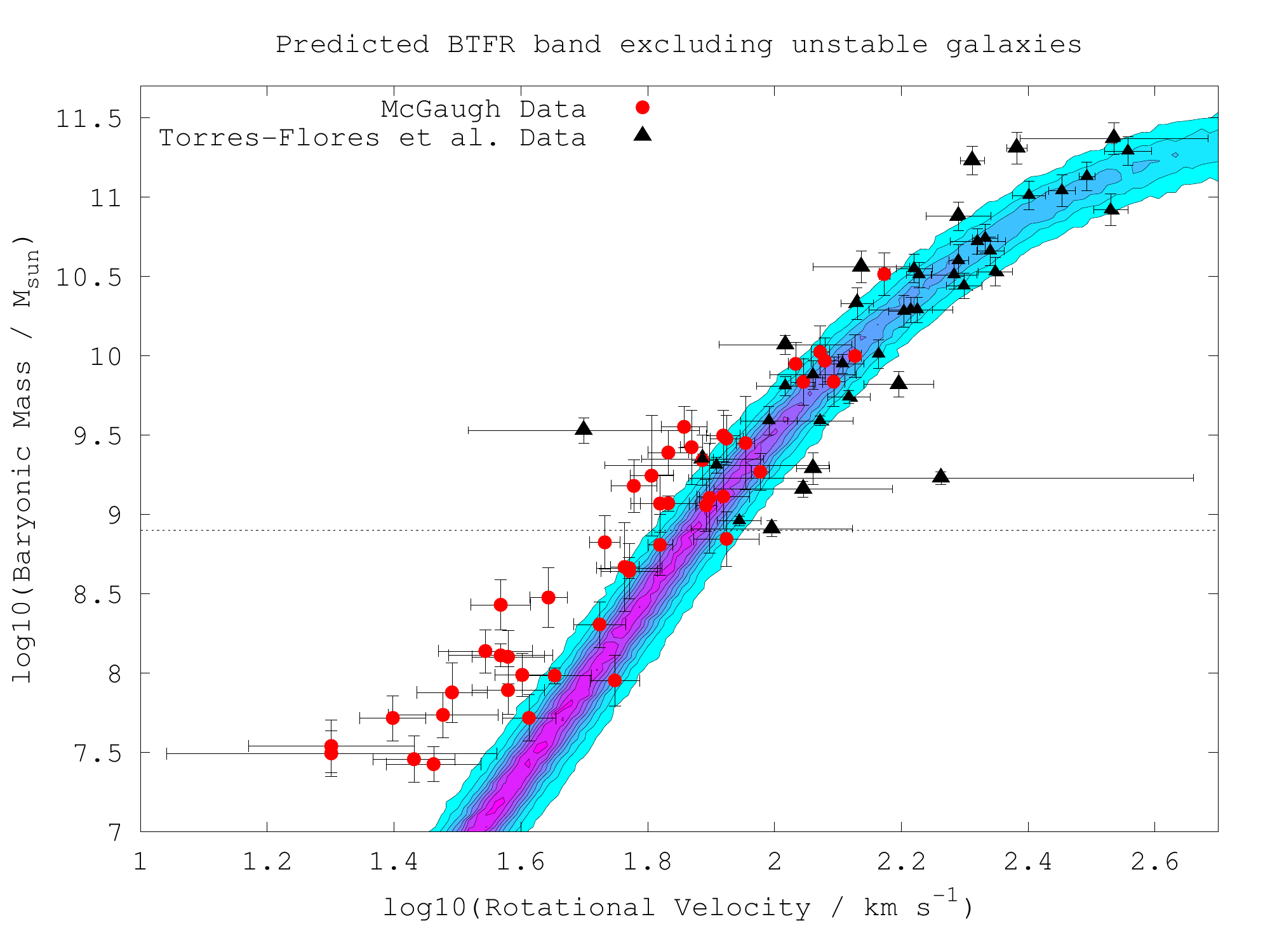}
  \caption{As Fig.~\ref{fig:Scatter}, but excluding model galaxies with $V_\mathrm{max} / \sqrt{GM_\mathrm{bar}/R_d} < 1.2$. See text and M98 (Sect. 3.2) for further details.}
  \label{fig:Scatter_Stability}
\end{figure}

Of the $\sim$250\,000 model galaxies, only 5\,169 are rejected using the stability threshold $e_{m,crit}$ = 1, 10\,475 for $e_{m,crit}$ = 1.1 and 17\,512 adopting the the upper limit deemed plausible by M98, $e_{m,crit}$ = 1.2 (a larger value causes rejection of halos with larger spins; see M98 eqs. 35 and 37). Thus we expect that the effect on the BTFR of rejecting galaxies unstable by this criterion is small, indicating that most of the model galaxies are in fact stable. This is illustrated in Fig.~\ref{fig:Scatter_Stability}, a replica of Fig.~\ref{fig:Scatter} but with galaxies unstable to bar formation at the $e_{m,crit}$ = 1.2 level removed. The majority of unstable galaxies have baryonic masses in the range $10^{10}$ -- $10^{10.5} M_{\odot}$, and the theoretical intrinsic scatter in this region is minorly reduced by excluding them. At lower-masses, where disagreement with the data is most pronounced, there is virtually no effect.

An important observational issue is the point on a galaxy's rotation curve at which velocity is measured. MG12 adopts $V_\mathrm{flat}$ (due to earlier work suggesting that this minimizes the scatter in the TFR), and defines a rotation curve as flat if the difference between the velocity at 3 disk scale-lengths and at the last measured point of the rotation curve is less than 15\%~\citep{McGaugh_Data_1}. This requirement is applied to many galaxies in the gas-rich data set, which are therefore not directly comparable to the data points for model galaxies with parameter values such that their rotation curves do not plateau. Here I apply a somewhat modified requirement to remove such galaxies, which should nevertheless produce similar results: the difference between the maximum velocity and that at 3 disk scale lengths must be less than 15\%.\footnote{The reason why McGaugh's flatness criterion cannot be applied directly is that the ``last measured point'' is not defined for the model galaxies. To improve the theoretical flatness requirement, one would require a radius marking the end of the observationally-resolved rotation curve. This radius would depend on the observational techniques and instruments used in the galaxy survey under consideration.} Only 200 model galaxies are excluded for having non-flat rotation curves, resulting in a negligible change to the BTFR band. Since rotation curves are typically observed to be flat in the outer regions of spiral galaxies, this gives some confidence that the galaxies produced by the methods and relations of Sect.~\ref{sec:Method} are realistic. These results also support the claim of~\citet{McGaugh_Complete} and~\citet{McGaugh_Data_1} that $V_\mathrm{flat}$ and $V_\mathrm{max}$ are typically very similar.

\section{Comparison with the literature}
\label{sec:Comp}

In this section I relate my results to those of TG11, D07 and MG12.

\subsection {Trujillo-Gomez et al. (2011)}
\label{sec:TG_Comp}

TG11 use the Bolshoi N-body simulation in association with baryon mass fractions from HAM to give a prediction for the BTFR which exhibits considerable curvature (their fig. 11). This presumably derives from the precise $M_\mathrm{vir}$-$M_\mathrm{bar}$ relation that is adopted (although this is not actually visible in TG11). The fact that no such curvature is evident in the data is pointed out in MG12, where it is used to quickly dismiss TG11's prediction; yet TG11 claim to find a ``good fit'' to the observational data. The extent to which curvature militates against a predicted BTFR depends on the status that the relation has within the paradigm under consideration. For TG11, the BTFR is simply an observed empirical relation that arises in some complicated way via galaxy formation, growth and merger history in addition to baryonic processes. The result is that, from this perspective, there is no reason for the BTFR to be linear in log-log space, and it is a coincidence if it approximately is. As long as the curved prediction is more or less consistent with each data point, the agreement may be considered adequate. The situation is very different, however, for McGaugh, who holds the BTFR to be a relation arising directly from a fundamental force law (MOND). From this persepective, the perfect linearity of the BTFR is one its defining characteristics, and hence inability to reproduce this is a major shortcoming. Furthermore, it can be seen in Fig.~\ref{fig:Lines} (or by comparison of MG12's fig. 5 with TG11's fig. 11) that although a power-law fit to both MG12's and TG11's data sets would have a best-fit slope of around 4, the intercept in the TG11 data is significantly lower. This increases agreement with their prediction.

Aside from the issue of whether or not the linearity of the BTFR is one of its fundamental properties, it is clear from fig. 11 in TG11 that the data points do lie above the predicted line at both low and high masses: the curvature does appear to be too great. This may be partially explained with the help of Fig.~\ref{fig:Lines}, in which the magenta line shows the analogue of the theoretical BTFR for elliptical galaxies. Contamination of TG11's HAM $M_\mathrm{vir}$-$M_\mathrm{star}$ relation with ellipticals would be expected to give a result between the red and magenta lines, which might exhibit considerable curvature.\footnote{However, the fact that spirals are far more abundant than ellipticals at low masses suggests that the HAM $M_\mathrm{vir}$-$M_\mathrm{bar}$ relation \emph{for all galaxies} -- as used by TG11 -- should be much closer to the spiral line than the elliptical line.} One might also expect to be able to explain the behaviour at the high-mass end of fig. 11 in this way. TG11's prediction lies below the data points describing spirals but is in agreement with those describing ellipticals. However, my Fig.~\ref{fig:Lines} shows that exchanging the expected spiral baryon mass fractions with those for ellipticals has very little effect on the predicted BTFR at high mass (if anything, the relation for ellipticals lies \emph{above} that for spirals). A possible explanation is that not only does the $M_\mathrm{vir}$-$M_\mathrm{bar}$ relation depend on galaxy morphology, but other relations important for calculation of the BTFR do too (e.g. the $M_\mathrm{vir}$-concentration relation). This reinforces the point made in Sect.~\ref{sec:Ellipticals} that the differences between the fundamental parameters of spirals and ellipticals are not sufficiently well-known to permit a fair comparison between them in baryonic mass-rotational velocity space.\footnote{Of course, the theoretical difference between spirals and ellipticals in this space is known only to the extent that the halos hosting spirals and ellipticals may be accurately distinguished (as described in RP11, Sect. 2.2.1). Methods for achieving this are still in their infancy and may be significantly refined in the future.}

A further difference between the Abundance Matching of TG11 and RP11 is that the former uses only the galaxies in their TFR sample, whilst the latter matches the theoretical halo mass function to the entirety of the Sloan Digital Sky Survey. Thus the $M_\mathrm{vir}$-$M_\mathrm{star}$ relation used by TG11 is built from around 1\,000 data points, whilst that described in RP11 contains of order 500\,000 and should therefore enjoy greater precision. Underneath this difference runs the question of whether the BTFR is a fundamental relation applicable to spirals of any mass, composition and location (a view championed by McGaugh), or whether it is a product of complex evolution histories in a $\Lambda$CDM universe. In the latter scenario, the BTFR may well be different for different sets of galaxies -- which may have been subject to significantly different conditions during their evolution -- and hence it would be reasonable to limit the $M_\mathrm{vir}$-$M_\mathrm{bar}$ relation to those galaxies used in the TFR. However, if all types of spiral lie on the same BTFR, it is unsatisfactory to ``explain'' the BTFR for only a small subset of these. A universal $M_\mathrm{vir}$-$M_\mathrm{bar}$ relation (as approximated for example by the results of RP11) is necessary to give the theoretical BTFR universal applicability.

\subsection {Dutton et al. (2007)}
\label{sec:D_Comp}

D07 use a semi-analytic approach, also based on M98, to investigate the LTFR in the $\Lambda$CDM paradigm, using simple parametrizations of the required relations which include tunable parameters. These are selected to fit the size-lumunosity (RL) relation in addition to the LTFR. D07 find that the slope and intercept of the $M_\mathrm{vir}$-$M_\mathrm{bar}$ relation can be tuned to fit the RL relation without significantly affecting the TFR. While it may be true that baryon mass fractions are of little importance to the luminosity TFR (the primary concern of D07), they are clearly of great importance to the baryonic TFR. For example, the BTFR described by the black line in Fig.~\ref{fig:Lines} is significantly steeper than early predictions for the BTFR adopting fixed baryon fractions (eg.~\citealt{MMW,Bullock}) primarily because of the steep $M_\mathrm{vir}$-$M_\mathrm{bar}$ relation predicted by HAM. Thus we see that successful prediction of the BTFR does not necessarily follow from successful prediction of the LTFR (cf. Sect.~\ref{sec:TFRs}).

In a similar vein, D07 claim that scatter in baryon mass fraction does not significantly affect the scatter in the LTFR, and, in view of the over-budgeted scatter in the RL relation, they set this to zero. Not only does this lack physical motivation (if the low observed baryon mass fractions are due to expulsion of baryons by astrophysical phenomena, one would clearly expect statistical fluctuations between galaxies), but it again neglects the fact that scatter in baryonic mass has a large effect on the BTFR. Indeed, this contributes a significant fraction of the scatter in Fig.~\ref{fig:Scatter}. The results in D07 are specific to the luminosity TFR, and many of the conclusions are unlikely to hold if the BTFR were considered also.

Finally, D07 argue that adiabatic contraction must be replaced by \emph{expansion} in order to successfully reproduce the LTFR and galaxy luminosity function with a realistic Initial Mass Function. Although I make no claim to ``successfully reproduce'' the BTFR data with my model, I do notice from Fig.~\ref{fig:Lines} that the effect of adiabatic contraction is very small and does, if anything, increase agreement with the data. D07's result may be specific to the parametrizations and parameter values produced by their particular optimisation procedure.


\subsection {McGaugh (2012)}
\label{sec:MG_Comp}

McGaugh argues that the failure of $\Lambda$CDM models to reproduce the low intrinsic scatter and high slope ($\sim$4) of a power-law fit to the BTFR without excessive fine tuning constitutes evidence against the paradigm. Given the existence of a Modified Gravity Model (MOND) which naturally explains these features, McGaugh further claims that the BTFR militates against the existence of dark matter itself. In this subsection I discuss the way in which my results impinge on two aspects of this argument.

In MG12, McGaugh claims that natural $\Lambda$CDM predictions have a slope of approximately 3. This was indeed the case for the ``na\"{\i}ve'' derivation of Sect.~\ref{sec:Prelim} in addition to several subsequent studies that introduced a dependence of halo concentration on $M_\mathrm{vir}$~\citep[e.g.][]{Bullock}. However, the slope is significantly steepened by replacing the constant baryon mass fractions used in these works by more ``realistic'' values, such as those derived from Halo Abundance Matching. For example, the low-mass end of the black line in Fig.~\ref{fig:Lines} has a slope in excess of 5. On the one hand, the HAM $M_\mathrm{vir}$-$M_\mathrm{bar}$ relation could be said to be a ``natural'' $\Lambda$CDM prediction in the sense that the halo mass function used in the Abundance Matching is determined from N-body simulations of dark halos in a $\Lambda$CDM cosmological background. On the other hand, a theoretical understanding of the way in which the appropriate number of baryons is ejected during galaxy formation (and where they went) remains wanting~\citep[e.g.~see][]{McGaugh_Baryons_1,McGaugh_Baryons_2}. This is a more fundamental problem for $\Lambda$CDM than failure to accurately reproduce the observed BTFR. Indeed, it is possible that the remaining discrepency between the observed and predicted BTFRs in Figs.~\ref{fig:Lines} and~\ref{fig:Scatter} is entirely attributable to errors in the adopted $M_\mathrm{vir}$-$M_\mathrm{bar}$ relation. It is clear that this relation makes a huge difference to the BTFR (contrast my results with those of TG11 and~\citealt{Bullock}), yet the method of HAM is only just now being fully developed, and may be subject to major refinements in the future. We also see from Fig.~\ref{fig:Scatter} that the BTFR prediction is significantly different to the data only in the region in which the HAM $M_\mathrm{vir}$-$M_\mathrm{bar}$ relation must be extrapolated. Could this be due to the form of this relation changing at low halo masses?

McGaugh claims that the scatter of gas-rich galaxies around a power-law BTFR may be fully accounted for by the uncertainties on the data points. Even if this were so,\footnote{Significant controversy surrounds this issue. For example,~\citet{Foreman_Scott} find McGaugh's gas rich data to be inconsistent with any zero-intrinsic scatter relation at the 95\% confidence level, and many other observational studies of the BTFR find at least some evidence for a non-zero intrinsic scatter.} it would not preclude the possibility of a non-zero intrinsic scatter in the underlying theoretical relation. Many potential BTFR predictions, with a range of non-zero intrinsic scatters, may be ``compatible'' with the data in the sense that they cannot be rejected at the 95\% confidence level. McGaugh's argument succeeds in demonstrating that the data is consistent with MOND (which no doubt is worthy of consideration for providing a simple explanation of this and many other aspects of galaxy phenomenology), but not that it is inconsistent with $\Lambda$CDM. Indeed, the magnitude of the theoretical scatter in Fig.~\ref{fig:Scatter} does not appear wildly discrepant with that in the data.

\section{Summary and suggestions for further work}
\label{sec:Conclusion}

This paper illustrates methodically the way in which the ``na\"{\i}ve'' slope-3 $\Lambda$CDM prediction for the BTFR can be improved by taking proper account of concentration, baryon mass fraction, and baryon disk collapse followed by adiabatic halo contraction. The slope of the relation is significantly increased by adopting baryon mass fractions from spiral-only Halo Abundance Matching, improving agreement with observational data at moderate to high baryonic masses. Predicted rotational velocities are systematically higher than those observed at the low-mass end, although plotting the theoretical BTFR in this region requires extrapolation of the $M_\mathrm{vir}$-$M_\mathrm{bar}$ relation obtained by the Abundance Matching. In addition, curvature (in log-log space) is introduced into the prediction and becomes significant for $M_\mathrm{bar} \gtrsim 10^{11} M_{\odot}$. Although curvature is not evident in the data, spirals with such a high baryonic mass are rare, and the observational studies used here do not preclude a reduction in the slope of the BTFR at very high masses. The effect of adiabatic contraction is small, especially at low and very high baryonic masses. The BTFR prediction obtained here is reasonably robust against changes to the density profile used to describe the dark matter halos, and the angular momentum fraction posessed by the baryons.

Accounting for the expected variability of the halo spin parameter, concentration, and baryon mass fraction, I find the theoretical BTFR to have an intrinsic scatter that is not clearly discrepant with that in the data, although may be somewhat larger than necessary to successfully account for the observations. Two simple selection requirements imposed on the model galaxies -- a flat rotation curve and stability against bar formation -- were found to have minimal effect on the predicted BTFR. This suggests that the spirals formed in the $\Lambda$CDM model presented here have rotation curves that plateau (as typically observed) and are predominantly stable.

There are two main ways in which this study could be usefully supplemented.

\begin{enumerate}

\item Only a subset of the effects causing intrinsic scatter in the TFR were considered here. Further sources of scatter include the possibility that dark halos have yet to virialise or are not well-fitted by an NFW profile~\citep{Jing}, variations in galaxies' mass aggregation histories~\citep{Eisenstein}, and halo triaxiality~\citep{Triaxiality}. However, there exist also unconsidered selection factors which will reduce the scatter, such as the possibility that disk formation in halos with high baryon mass fraction is likely to create elliptical or S0 galaxies to which the TFR does not apply~\citep{Mayer_Moore}. The true theoretical intrinsic scatter can only be determined once all of these effects have been included in the style of Sect.~\ref{sec:Selection}. Proper implementation of these selection criteria is contingent on a solid theoretical understanding of galaxy formation.

\item The comparison with observational data presented in Sect.~\ref{sec:Results} is far from optimal. As discussed in Sect.~\ref{sec:Data}, the ideal would be to amalgamate all data sets that employ compatible mass-to-light ratios, prescriptions for gas mass measurement, methods of rotational velocity measurement and selection criteria. By implementing similar constraints in the theoretical model, a comparison could be made between theory and observation that would be free from systematic errors. A simple hypothesis test could then be used to give a quantitative measure of the agreement between the BTFR and the $\Lambda$CDM paradigm as we currently understand it.

\end{enumerate}

The results of this study will be modified by advances in the techniques of Halo Abundance Matching and halo spin and concentration estimation from N-body simulations. The modular approach will make such updates easy to implement, and their effects on the BTFR readily visible.

\begin{acknowledgements}

I would like to thank Subir Sarkar for guidance and encouragement, and Stacy McGaugh for helpful discussions of his work and mine.

\end{acknowledgements}

\bibliographystyle{aa}
\bibliography{Desmond_baryonic_Tully_Fisher_Relation_in_LCDM}

\end{document}